\documentclass[aps,pra,twocolumn,superscriptaddress]{revtex4}
\usepackage{times,amsfonts}
\usepackage{amsmath}	
\usepackage{amssymb}
\usepackage[ansinew]{inputenc}
\usepackage{cancel}
\usepackage{graphicx}

\usepackage{color}
\usepackage[normalem]{ulem}
\usepackage{comment}

\DeclareMathOperator*{\Simiq}{\simeq}

\newcommand{\ve}[1]{{\mathbf #1}}
\newcommand{\Frac}[2]{\displaystyle\frac{#1}{#2}}

\begin{document}

\title{Phase Equilibrium of Binary Mixtures in Mixed Dimensions}
	
\author{E. Malatsetxebarria} 
\affiliation{Centro de F\'isica de Materiales CSIC-UPV/EHU and
  Donostia International Physics Center (DIPC), Paseo Manuel de
  Lardizabal, E-20018 San Sebasti\'an, Spain}

\author{F. M. Marchetti}
\affiliation{Departamento de F\'isica Te\'orica de la Materia
  Condensada, Universidad Aut\'onoma de Madrid, Madrid 28049, Spain}

\author{M. A. Cazalilla}
\affiliation{Centro de F\'isica de Materiales CSIC-UPV/EHU and
  Donostia International Physics Center (DIPC), Paseo Manuel de
  Lardizabal, E-20018 San Sebasti\'an, Spain}
\affiliation{Graphene Research Centre
National University of Singapore,
6 Science Drive 2,
Singapore 117546.}

\date{April 23, 2013}

\begin{abstract}
  We study the stability of a Bose-Fermi system loaded into an array
  of coupled one-dimensional (1D) ``tubes'', where bosons and fermions
  experience different dimensions: Bosons are heavy and strongly
  localized in the 1D tubes, whereas fermions are light and can hop
  between the tubes. Using the $^{174}$Yb-$^{6}$Li system as a
  reference, we obtain the equilibrium phase diagram. We find that,
  for both attractive and repulsive interspecies interaction, the
  exact treatment of 1D bosons via the Bethe ansatz implies that the
  transitions between pure fermion and any phase with a finite density
  of bosons can only be first order and never continuous, resulting in
  phase separation in density space. In contrast, the order of the
  transition between the pure boson and the mixed phase can either be
  second or first order depending on whether fermions are allowed to
  hop between the tubes or they also are strictly confined in 1D. We
  discuss the implications of our findings for current experiments on
  $^{174}$Yb-$^{6}$Li mixtures as well as Fermi-Fermi mixtures of
  light and heavy atoms in a mixed dimensional optical lattice system.
\end{abstract}








\maketitle

\section{Introduction}
\label{sec:intro}
The quest for lower temperatures in ultracold gases has lead to the
development of many ingenious techniques to cool several atomic and
molecular species. In particular, the explosion of activity concerning
ultracold Fermi gases has become possible largely owing to the success
of sympathetic cooling, which allows to efficiently cool fermions by
mixing them with bosons~\cite{pethicsmith}. At the same time, this
procedure has stimulated the investigation, both experimental and
theoretical, of Bose-Fermi mixtures. Here, the possibility of tuning
the inter-species interaction strength using Feshbach
resonances~\cite{PhysRevLett.81.69,Inouye:1998fk, Bloch:2008gl}, has
led to the exploration of many interesting phenomena such as collapse
and phase separation~\cite{PhysRevLett.80.1804,PhysRevA.61.053605}, as
well as boson-mediated Cooper pairing~\cite{pethicsmith}. Furthermore,
Feshbach resonances can also be used to generate heteronuclear
molecules, which can exhibit large electric dipole moments. This opens
the interesting possibility of studying dipole-dipole interactions in
quantum degenerate
gases~\cite{PhysRevLett.97.180413,PhysRevLett.98.060403,PhysRevA.80.053610,PhysRevLett.85.1791,Micheli:2006wy,
  Buchler:2007fo,DeMille:2002wq}.

Meanwhile, the advent of optical-lattice
confinement~\cite{greiner2002,Bloch:2008gl} has turned ultracold
atomic gases into unique environments where to simulate and understand
strongly correlated phenomena relevant to condensed matter systems.
This has been made it possible, for example, by confining the atomic
clouds in reduced dimensions, such as a one dimensional (1D) array of
two-dimensional (2D) planes or a 2D array of 1D tubes. Whereas the
former has enabled the study of interesting phenomena occurring in
two-dimensions~\cite{Posazhennikova:2006kp,Martiyanov:2010kt}, the
latter has provided us with an amazingly tunable tool to explore the
physics of interacting 1D quantum
systems~\cite{Moritz:2005bf,Kinoshita:2004jp,paredesnature,RMP_Cazalilla},
of which it is much more difficult to find faithful realizations in a
more conventional condensed matter context.
\begin{figure}[hb]
\includegraphics[width=1.0\linewidth,angle=0]{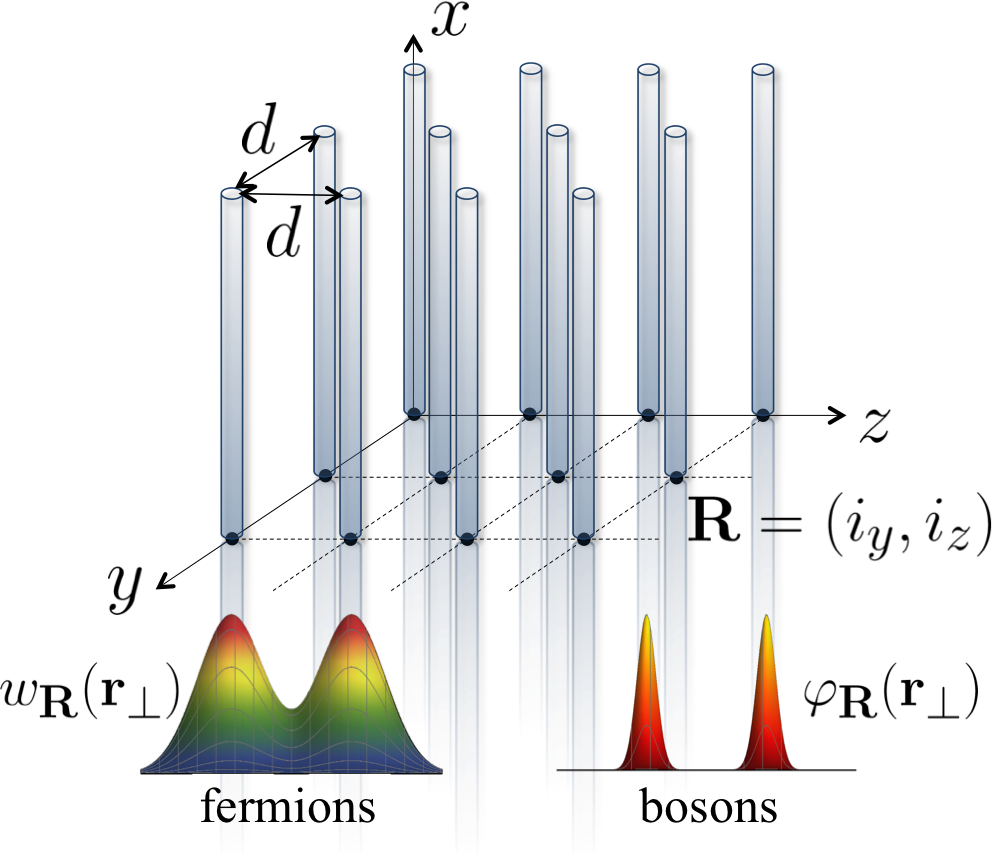}
\caption{(Color online) Schematic representation of the system studied
  in this work, namely a mixed dimensionality Bose-Fermi system. The
  Bose-Fermi cloud is loaded in an anisotropic optical lattice that
  produces a two-dimensional $N_y \times N_z$ array of one-dimensional
  (1D) tubes of length $L$.  The fermions are light enough to hop
  between the tubes whereas bosons, which are assumed to be much
  heavier, are confined in the 1D tubes. The bosonic and fermionic
  Wannier functions ($\varphi_{\ve{R}} (\ve{r}_{\perp})$ and
  $w_{\ve{R}} (\ve{r}_{\perp})$, respectively) are also displayed.}
\label{fig:schem}
\end{figure}

Optical lattice confinement has also allowed to envisage the
realization of new types of quantum systems.  One such example,
analyzed in this work, is provided by mixtures of interacting
particles in mixed-dimensional lattices. In recent years, these
systems have attracted an increasing amount of theoretical
attention~\cite{PhysRevLett.101.170401,Yang:2011jc}, and very recently
they have been also experimentally realized~\cite{Lamporesi:2010eb,
  Minardi:2011eb}.  Besides its intrinsic interest as a new category
of quantum many-body systems, they may also offer advantages for
reducing few-body losses and enhancing stability in strongly
interacting regimes~\cite{Marchetti:2009de}.
 
In recent years, there has been a growing interest in understanding
the properties of mixtures of ultracold Bose and Fermi gases. In
particular, binary mixtures of bosons and spin-polarized fermions have
been studied in three-dimensions (3D)~\cite{Albus:2003dy,
  PhysRevLett.80.1804, PhysRevA.61.053605, Viverit:2002hg,
  Modugno:2003fw,Gunter:2006bj}, 2D~\cite{Buchler:2004fn} and
1D~\cite{Das:2003jr,PhysRevA.3.393, Cazalilla:2003gj,
  Mathey:2004jb,Imambekov:2006ib}. Note that the 1D geometry has a
special relevance because of the central role played by quantum
fluctuations and the fact that there is neither broken continuous
symmetry nor, consequently, off-diagonal long-range order. The
equilibrium phase diagram of 1D Bose-Fermi mixtures has been
considered by many authors~\cite{PhysRevA.3.393,Das:2003jr,
  Cazalilla:2003gj, Mathey:2004jb,Imambekov:2006ib}, while the case of
a Bose-Fermi system in an anisotropic optical lattices was studied by
one of us in Ref.~\cite{Marchetti:2009de}.

In addition to dimensionality issues, the significant interest in
studying the stability of binary mixtures comes also from the
possibility of tuning the Bose-Fermi scattering length via the
Feshbach resonance mechanism. Indeed, in the repulsive interaction
regime, the spatial overlap between bosons and spin-polarized fermions
is reduced, thus ensuring the stability of the
system~\cite{Zaccanti:2006bf, 2007cond.mat..2277M}. When the repulsion
is increased, the two components tend instead to phase separate,
rather than uniformly mix: In the particular case of a 3D geometry,
phase separation occurs either between a mixed phase and a purely
fermionic phase or between two pure
phases~\cite{PhysRevA.61.053605}. In the regime where the interaction
between bosons and fermions is attractive, a significant reduction of
the interatomic distance can lead to a collapse of the mixture,
because of three-body recombination processes~\cite{Zirbel:2008hl}. As
discussed later, the stability of the mixture towards a collapsed
phase can be enhanced in one-dimensional
geometries~\cite{Marchetti:2009de}.

In this article, we study the phase stability of a Bose-Fermi mixture
embedded in a mixed-dimensional optical lattice of an array of
one-dimensional tubes~\cite{PhysRevLett.81.3108, Stoferle:2004ji} (see
Fig.~\ref{fig:schem}).  The mixed-dimensionality comes from the fact
that, while bosons are longitudinally confined along the tubes and
strictly move in 1D, fermions are not constrained to 1D and are
allowed to hop between tubes in the transverse directions. This
assumption can be justified based on the fact that many realizations
of Bose-Fermi mixtures consider bosonic species that are much heavier
than the fermionic ones. We do also assume that, while bosons interact
with each other, as well as with fermions, the fermionic component is
polarized in a single hyperfine state and thus is non-interacting at
very low temperatures.
 
As already mentioned, one motivation to study this geometry is that,
by confining bosons in 1D, the mixture stability is
enhanced~\cite{Marchetti:2009de}. Note that, even for purely bosonic
gases, the spatial overlap between bosons has been experimentally
shown to be strongly suppressed by the strong correlations emerging in
1D~\cite{kinoshita05}. In addition, mixed-dimensional systems allow to
study interesting few-body~\cite{2012PhRvL.109b0401T} and
many-body~\cite{Yang:2011jc,dissip} phenomena. We should stress that
the results reported here are relevant to several experimental
realizations. The simplest is, for example, a mixture of light
fermions and heavy bosons. In particular, in order to make contact
with ongoing and future experiments~\cite{Hansen:2013gp,
  2011PhRvL.106o3201I, 2010ApPhB..98..691O}, we explicitly consider a
mixture of $^6$Li (a light fermion) and $^{173}$Yb (a heavy boson)
atoms.  An alternative realization could be given by an originally
imbalanced mixture of fermions in two hyperfine
states~\cite{Chin:2006dt}: Bosons are then formed by associating
fermions into Feshbach molecules~\cite{Chin:2006dt}, leaving out from
the pairing the spin polarized excess majority fermions. When the
fermions belong to the same atomic species, the Feshbach molecules
have twice the mass and twice the polarizability of the fermionic
atoms. Therefore, it should be relatively easy to confine the bosons
in 1D by loading them into a two-dimensional optical lattice. On the
same lattice, the remaining majority fermions behave as the light
component.

We find that, since bosons are confined to strictly 1D, they can
undergo fermionization. This means that, as they mix with fermions,
they form a Tonks-Girardeau gas whose energy per unit length grows as
third power of the lineal density. As a result, we find that the
nature of the transition from the pure Fermi gas to a mixture is
always first order, implying phase separation in density space.  Note
that this is a very different result from the one obtained by assuming
that bosons form a (quasi-)condensate, with the energy density growing
as the square of the boson lineal density. The latter situation
applies either to a high density Bose gas in 1D or to a gas of bosons
that can hop in 3D.  We also find that, in the mixed dimensionality
lattice, the transition between pure boson and mixed phases is
continuous, while it becomes first order when the fermions are also
confined in 1D.

The rest of this article is organized as follows: In
Sec.~\ref{sec:model} we introduce the model for the Bose-Fermi mixture
and discuss methods and approximations employed to derive the system
phase diagram. In particular, in Sec.~\ref{sec:bosef}, we explain how
a mean-field approximation is applied solely to the boson-fermion
interaction, whereas the boson-boson interaction is treated
non-perturbatively using the Bethe ansatz in Sec.~\ref{sec:freen}. In
Sec.~\ref{sec:phase}, we derive the phase diagram and interpret the
origin of the discontinuous character of the transitions between the
pure fermion and mixed phases in terms of an expansion of the free
energy for small values of the boson density. We first describe the
results obtained for the case of mixed dimensions in
Sec.~\ref{sec:mixdi} and later for the pure 1D limit in
Sec.~\ref{sec:1Dlim}. Finally, in Sec.~\ref{sec:concl}, we present the
main conclusions of our study and discuss the limitations of our
approach, as well as some directions for future work. Some technical
aspects of our derivations are provided in the appendices.

\section{Model}
\label{sec:model}
We consider a mixture of interacting bosonic ($B$) and
single-component fermionic ($F$) atoms described by the following
Hamiltonian (in $\hbar=1$ units):
\begin{align}
\label{eq:model}
  \hat{H}&=\hat{H}_B + \hat{H}_F + \hat{H}_{BF}\; ,\\ 
  \nonumber \hat{H}_{B}&= \int d\ve{r} \hat{\Psi}_B^{\dag}
  \left[\Frac{-\nabla^2}{2m_B} + V_B(\ve{r}) - \mu_{B} +
    \frac{g_{BB}}{2} \hat{\rho}_B\right] \hat{\Psi}^{\phantom{\dag}}_B
  \; ,\\
  \nonumber \hat{H}_{F}&= \int d\ve{r}\hat{\Psi}_F^{\dag}
  \left[\Frac{-\nabla^2}{2m_F} + V_F(\ve{r}) - \mu_F\right]
  \hat{\Psi}^{\phantom{\dag}}_F\; ,\\
  \nonumber \hat{H}_{BF}&= g_{BF}\int d\ve{r} \hat{\rho}_B(\mathbf{r})
  \hat{\rho}_F (\mathbf{r})\; ,
\end{align}
where the density operators are $\hat{\rho}_{\alpha}(\ve{r}) =
\hat{\Psi}^{\dag}_{\alpha} (\ve{r})
\hat{\Psi}^{\phantom{\dag}}_{\alpha} (\ve{r})$, with $\alpha = B,F$.
We have approximated all interaction potentials by contact
interactions, which are parameterized by an $s$-wave scattering length
$a_{\alpha\beta}$:
\begin{equation}
  g_{\alpha\beta} = \Frac{4\pi a_{\alpha\beta}}{m_{\alpha\beta}} \; ,
\label{eq:scatt}
\end{equation}
being $m_{BB}=m_B$ and $m_{BF}= 2m_B m_F/(m_B+m_F)$. For thermodynamic
stability reasons, the bosons are assumed to repel each other (i.e.,
$a_{BB}>0$). The value of $a_{BB}$ can be tuned by, e.g., controlling
the strength of the transverse
confinement~\cite{PhysRevLett.81.938}. The sign of the Bose-Fermi
interaction strength is determined by $a_{BF}$, which can be
controlled independently from $a_{BB}$ by resorting to an
inter-species Feshbach resonance. Below we consider both the repulsive
($a_{BF}>0$) as well as the attractive ($a_{BF}<0$) case. At ultracold
temperatures, interactions between identical fermions can be safely
neglected.

The Bose-Fermi mixture is loaded into an anisotropic optical lattice
formed by a 2D $N_y \times N_z$ square array of 1D ``tubes'' of length
$L$ directed along the $x$-direction and equally spaced by a distance
$d$ (see Fig.~\ref{fig:schem}). This can be described by an optical
potential of the form $V_{\alpha}(\ve{r}_{\perp}) = V^{\alpha}_0
[\sin^2 (\pi y/d)+\sin^2 (\pi z/d)]$, where $\ve{r}_{\perp} =
(y,z)$. The strength of the optical potential $V^{B}_0$ ($V^{F}_0$) is
determined by both the laser intensity and the boson (fermion) atomic
polarizability, allowing the possibility of mixed dimensionality for
the mixture. In particular, we assume that the bosons are tightly
confined in 1D tubes and thus move strictly in 1D, while fermions can
hop between the tubes. We will derive the thermodynamic phase diagram
for this geometry, thus neglecting the harmonic confinement. By making
use of the local density approximation, information about the
experimentally relevant trapped case can be extracted from the
homogeneous phase diagram plotted in chemical potential space.

Because the bosons in the mixture are assumed to be more massive than
fermions and/or to experience a more confining lattice potential
$V_B(\ve{r}_{\perp})$, they are tightly confined along the ``tubes'',
a configuration often refereed to as a two-dimensional optical
lattice~\cite{greiner2002,Kinoshita:2004jp,
  Stoferle:2004ji,Moritz:2003kv,Cazalilla:2006kh,RMP_Cazalilla}.
Thus, the field operator $\hat{\Psi}^{\phantom{\dag}}_{B} (\ve{r})$
can be expressed in terms of Wannier functions
$\varphi^{\phantom{*}}_{\ve{R}} (\ve{r}_\perp)$ localized at the tube
site $\ve{R}=(i_y, i_z) d$ (see, e.g.,~\cite{Bloch:2008gl} and
references therein) and the tube boson operator
$\hat{\Psi}^{\phantom{\dag}}_{B\ve{R}} (x)$:
\begin{equation}
  \hat{\Psi}^{\phantom{\dag}}_{B} (\ve{r}) = \sum_{\ve{R}}
  \varphi^{\phantom{*}}_{\ve{R}} (\ve{r}_\perp)
  \hat{\Psi}^{\phantom{\dag}}_{B\ve{R}} (x)\; .
\end{equation}
The Wannier functions form an orthonormal basis. By neglecting the
interactions between tubes with $\mathbf{R} \neq \mathbf{R}^{\prime}$,
we can rewrite the bosonic Hamiltonian $\hat{H}_B$ in~\eqref{eq:model}
as a sum of decoupled 1D Hamiltonians,
\begin{multline}
  \hat{H}_{B}= \sum_{\mathbf{R}}\int dx \hat{\Psi}_{B\ve{R}}^\dag (x)
  \left[\Frac{-\partial_x^2}{2m_B} -\mu_{B}^{1D} \right.\\
  \left.+ \frac{g^{1D}_{BB}}{2} \hat{\rho}_{B\ve{R}}(x)\right]
  \hat{\Psi}^{\phantom{\dag}}_{B\ve{R}} (x)\; ,
\label{eq:boseh}
\end{multline}
where $\hat{\rho}_{B\ve{R}} (x) = \hat{\Psi}_{B\ve{R}}^{\dag} (x)
\hat{\Psi}^{\phantom{\dag}}_{B\ve{R}} (x)$ is the single-tube boson
density operator, $g^{1D}_{BB}$ is the one-dimensional boson coupling,
which, for weak boson-boson interaction, takes the form:
\begin{equation}
  g_{BB}^{1D}= g_{BB}\int d\mathbf{r}_\perp
  |\varphi^{\phantom{*}}_{\ve{R}} (\mathbf{r}_\perp)|^4\; ,
\label{eq:appro}
\end{equation}
and $\mu_{B}^{1D}$ is the 1D boson chemical potential:
\begin{equation*}
  \mu_{B}^{1D}= \int d\ve{r}_\perp \varphi_{\ve{R}}^* (\ve{r}_\perp)
  \left[\Frac{\nabla^2_{\perp}}{2m_B} - V_B (\ve{r}_{\perp})
    +\mu_{B} \right] \varphi^{\phantom{*}}_{\ve{R}}
  (\ve{r}_\perp) \; .
\end{equation*}
Note that, as we will see later, for strong boson-boson interactions,
the expression of the 1D boson coupling $g^{1D}_{BB}$ is instead given
by Eq.~\eqref{eq:olsha}~\cite{PhysRevLett.81.938} rather than
Eq.~\eqref{eq:appro}.

In contrast, we assume the fermions to be more weakly confined than
bosons along each tube due to their smaller mass and/or a weaker
optical potential. Yet, the lattice confinement is strong enough so
that the description of the Fermi field in terms on the lowest Bloch
band $\phi_{\ve{k}_{\perp}} (\ve{r}_{\perp})$ is accurate and we can
expand:
\begin{equation}
  \hat{\Psi}^{\phantom{\dag}}_{F} (\ve{r}) = \Frac{1}{\sqrt{L}}
  \sum_{\ve{k}} e^{ik_{x}x} \phi_{\ve{k}_{\perp}} (\ve{r}_{\perp})
  \hat{f}^{\phantom{\dag}}_{\ve{k}}\; ,
\label{eq:fexpa}
\end{equation}
where $\ve{k}=(k_x, \ve{k}_{\perp})$. Here, whereas the motion along
the $x$ direction is free, the motion in the transverse directions
$\ve{r}_{\perp} = (y,z)$ is described by $\phi_{\ve{k}_{\perp}}
(\ve{r}_{\perp})$, which is a Bloch wavefunction belonging to the
lowest Bloch band characterized by a crystal momentum
$\ve{k}_{\perp}$. Projecting the fermion Hamiltonian onto this band
yields:
\begin{equation}
  \hat{H}_{F} = \sum_{\ve{k}} \left[ \varepsilon(\ve{k}) - \mu_{F}
    \right] \hat{f}^{\dag}_{\ve{k}} \hat{f}^{\phantom{\dag}}_{\ve{k}}
  \; ,
\label{eq:fermh}
\end{equation}
 where the Fermion dispersion reads as:
\begin{align}
 \varepsilon(\ve{k}) &= \Frac{k^2_x}{2 m_F} + \epsilon(\ve{k}_\perp)\;
 \\
 \epsilon(\ve{k}_\perp) &= 2t \left[2 - \cos(k_{y}d) -\cos(k_{z}d)
   \right]\; .
\end{align}

Thus, summarizing, in the geometry studied here, the bosons are
tightly confined to move in 1D, whereas the fermions can hop between
the tubes, although the optical lattice potential does affect their
dispersion relation. Under these conditions, it is
known~\cite{RMP_Cazalilla} that, at low temperatures, the bosonic
atoms loose their individuality and the low-energy long-wavelength
excitations are 1D phonons.  For arbitrary values of $g^{1D}_{BB} >
0$, the ground state properties of such an interacting bosonic gas are
exactly described by the Bethe ansatz solution obtained by Lieb and
Liniger~\cite{Lieb:1963vu}. The lack of individuality and the highly
correlated behavior brought about by the 1D confinement calls for a
treatment of the problem that treats the boson-boson interactions in a
non-perturbative way. Since a mean-field approximation fails to
capture the fundamental bosonic correlations in 1D, we apply it only
to the interactions between the fermions and the bosons, as we explain
in what follows.

\subsection{Bose-Fermi interaction: mean-field}
\label{sec:bosef}
In order to render the above model tractable, we apply a mean-field
approximation to the Bose-Fermi interaction term. We do this in such a
way that the different 1D tubes become decoupled at the expense of
introducing self-consistent shifts of both the boson and fermion
chemical potentials. To this end, we first observe that the tight
confinement of the bosons in 1D allows us to neglect the overlap
between Wannier functions localized at different 1D tubes (see
Fig.~\ref{fig:schem}) and thus we can approximate the density operator
of the bosons as
\begin{equation*}
  \hat{\rho}_{B}(\ve{r}) \simeq \sum_{\ve{R}}
  |\varphi^{\phantom{*}}_{\ve{R}} (\ve{r}_\perp)|^2
  \hat{\rho}_{B\ve{R}} (x) \; .
\end{equation*}
In this limit, the Bose-Fermi interaction term $\hat{H}_{BF}$ of the
Hamiltonian~\eqref{eq:model} can be written as
\begin{equation}
  \hat{H}_{BF} \simeq g_{BF} \sum_{\ve{R}} \int dx
  \hat{\rho}_{B\ve{R}}(x) \hat{\rho}_{F\ve{R}}(x)\; ,
\label{eq:mfst1}
\end{equation}
where $\hat{\rho}_{F\ve{R}}(x)$ is a projection of the 3D Fermi
density operator $\hat{\rho}_F (\ve{r})$ on the $\ve{R}$-th tube:
\begin{equation}
  \hat{\rho}_{F\ve{R}}(x) = \int d\ve{r}_{\perp}
  |\varphi^{\phantom{*}}_{\ve{R}} (\ve{r}_{\perp})|^2 \hat{\rho}_F
  (\ve{r})\; .
\label{eq:fsing}
\end{equation}
We note that this approximation does not decouple the different tubes
yet because, even if bosons cannot hop from one tube to another, the
interaction between bosons belonging to different tubes is mediated by
the hopping fermions, i.e., the operator
$\hat{\rho}_{F\ve{R}}(x)$. However, if we rely on a mean-field
approximation to replace the operator $\hat{\rho}_{F\ve{R}}(x)$ by its
expectation value (which, as shown in detail in App.~\ref{app:avera},
is a constant), the different tubes become decoupled, which allows for
a solution of the model introduced above. We emphasize again that this
kind of mean-field approximation is different from the standard
treatment (for example employed in Ref.~\cite{Marchetti:2009de}),
where also the boson density in the boson-boson interaction term is
replaced with its expectation value.  Instead, here the boson
interaction is treated non-perturbatively using the Bethe ansatz,
emphasizing that the fundamental entities subject to the mean-field
interaction are the 1D tubes and not the bosons themselves.

Thus, in Eq.~\eqref{eq:mfst1}, we write the density operators
$\hat{\rho}_{\alpha \ve{R}}(x)$ as their quantum averages plus
fluctuations, i.e.  $ \hat{\rho}_{\alpha \ve{R}}(x) = \langle
\hat{\rho}_{\alpha \ve{R}}(x) \rangle + \delta\hat{\rho}_{\alpha
  \ve{R}}(x)$. Hence, the mean-field approximation is obtained by
substituting these expressions into Eq.~\eqref{eq:mfst1} and by
neglecting the second order terms in the fluctuations, which leads to:
\begin{multline}
  \hat{H}_{BF} \simeq H^{mf}_{BF} = g_{BF} \sum_{\ve{R}} \int dx
  \left[- \langle\hat{\rho}_{B\ve{R}}(x)\rangle \langle
    \hat{\rho}_{F\ve{R}}(x)\rangle \right.\\
    \left. + \hat{\rho}_{B\ve{R}}(x) \langle \hat{\rho}_{F\ve{R}} (x)
    \rangle + \langle \hat{\rho}_{B\ve{R}}(x) \rangle
    \hat{\rho}_{F\ve{R}} (x) \right] \; .
\label{eq:mfexp}
\end{multline}
In absence of harmonic confinement along the tubes, translational
invariance along the $x$-direction requires that the averages
\begin{align}
  \langle \hat{\rho}_{B\ve{R}}(x)\rangle &= \rho^{0}_{B} & \langle
  \hat{\rho}_{F\ve{R}}(x)\rangle &= A \: \rho^{0}_{F}
\label{eq:aver}
\end{align}
are constants independent on the tube index $\ve{R}$. Here, the constant 
\begin{equation}
  A\simeq N \int d\ve{r}_{\perp} |\varphi^{\phantom{*}}_{\ve{R}}
  (\ve{r}_{\perp})|^2 |\phi_{\ve{k}_{\perp}} (\ve{r}_{\perp})|^2\; ,
\label{eq:acons}
\end{equation}
it is obtained in the limit where the transverse confinement for the
bosons is tight (see App.~\ref{app:avera} for the details of the
derivation). Also, we have introduced the following \emph{lineal}
densities:
\begin{equation}
  \rho_{\alpha}^0 = \Frac{N_{\alpha}}{N L}\; ,
\label{eq:linde}
\end{equation}
where $N=N_y N_z$ is the total number of 1D tubes. Thus, within this
mean-field approximation, the system Hamiltonian~\eqref{eq:model} can
be written as
\begin{equation}
  \hat{H} \simeq \hat{H}^{mf}=\hat{H}_{B}^{mf} + \hat{H}_{F}^{mf} - \sum_{\ve{R}}
  g_{BF} A \rho^{0}_{B} \rho^{0}_{F}\; ,
\label{eq:meanf}
\end{equation}
where $\hat{H}_{B}^{mf}$ is defined as the bosonic Hamiltonian from
Eq.~\eqref{eq:boseh} with the chemical potential shifted as
$\mu_{B}^{1D} \to \mu_{B}^{1D} - g_{BF} A \rho^{0}_{F}$:
\begin{multline}
  \hat{H}_{B}^{mf} = \sum_{\ve{R}} \int dx \hat{\Psi}_{B\ve{R}}^\dag
  (x) \left[\Frac{-\partial_x^2}{2m_B} -\mu_{B}^{1D} + g_{BF} A
    \rho^{0}_{F} \right.\\
  \left.+ \frac{g^{1D}_{BB}}{2} \hat{\rho}_{B\ve{R}} (x)\right]
  \hat{\Psi}^{\phantom{\dag}}_{B\ve{R}} (x)\; ,
\label{eq:mfbos}
\end{multline}
and $\hat{H}_{F}^{mf}$ is the fermion Hamiltonian from
Eq.~\eqref{eq:fermh} with a chemical potential shifted as $\mu_{F} \to
\mu_{F} - g_{BF} A \rho^{0}_{B}$:
\begin{equation}
  \hat{H}_{F}^{mf} = \sum_{\ve{k}} \left[ \varepsilon (\ve{k}) -
    \mu_{F} + g_{BF} A \rho^{0}_{B} \right] \hat{f}^{\dag}_{\ve{k}}
  \hat{f}^{\phantom{\dag}}_{\ve{k}}
\label{eq:fermf}
\end{equation}
We would like to stress that we are not applying a mean-field
approximation to the boson-boson interaction term $\hat{\rho}^2_{B
  \ve{R}}(x)$. On the contrary, as shown in the next section, we shall
treat this term exactly using the Bethe ansatz solution of
Eq.~\eqref{eq:mfbos} due to Lieb and Liniger~\cite{Lieb:1963vu}.

\subsection{Zero-temperature free energy}
\label{sec:freen}
Starting from the mean-field Hamiltonian defined by
equations~\eqref{eq:meanf},~\eqref{eq:mfbos}, and~\eqref{eq:fermf}, we
can evaluate the grand-canonical free energy density at zero
temperature. Note that, by virtue of the mean-field approximation and
the translational invariance, the bosonic
$\hat{\Psi}^{\phantom{\dag}}_{B\ve{R}}$ and fermionic
$\hat{f}^{\phantom{\dag}}_{\ve{k}}$ field operators in
Eq.~\eqref{eq:meanf} have become decoupled and therefore we can
separate the bosonic and a fermionic contributions to the free energy
potential $f=f(\mu_B^{1D}, \mu_F, \rho_B^0)$, which can be written as:
\begin{equation}
  f = \frac{(\rho_B^0)^3}{2m_B} e(\gamma) - \mu_B^{1D} \rho_B^0
  -\Frac{1}{N} \sum_{\ve{k}_{\perp} \in BZ} \Frac{k_{Fx}^3
    (\ve{k}_{\perp})}{3 \pi m_F} \; ,
\label{eq:freen}
\end{equation}
where $BZ$ stands for the 2D Brillouin zone, i.e., the region of
$\ve{k}_{\perp}$-space where $|k_{y,z}| \le \pi/d$. In evaluating this
expression, we have relied upon the Bethe ansatz
solution~\cite{Lieb:1963vu,RMP_Cazalilla} of the interacting 1D
boson Hamiltonian, $\hat{H}_{B}^{mf}$~\eqref{eq:mfbos}. It is worth
noting that the constant term $g_{BF} A \rho^{0}_{B} \rho^{0}_{F}$ in
Eq.~\eqref{eq:meanf} cancels exactly the mean-field shift of the
bosonic chemical potential, $\mu_{B}^{1D} \to \mu_{B}^{1D} -
g_{BF} A \rho^{0}_{F}$, in Eq.~\eqref{eq:mfbos}. 

In Eq.~\eqref{eq:freen}, the dimensionless function $e(\gamma)$, where
$\gamma = m_B g_{BB}^{1D}/\rho_B^0$, is determined by numerically
solving the following system of coupled integral equations:
\begin{align}
  e(\gamma) &= \frac{\gamma^3}{l^3} \int_{-1}^{1} du u^2 g(u)\; ,\\
  2\pi g(u) &= 1+ 2l \int_{-1}^{1} du' \frac{g(u')}{(u-u')^2 + l^2}\; ,
\end{align}
where $l = \gamma \int_{-1}^1   du\:  g(u)$.  The fermionic contribution to
the free energy can be expressed as an integral over the transverse
momentum, which leads to the last term in Eq.~\eqref{eq:freen}, where
we have defined:
\begin{equation*}
  k_{Fx} (\ve{k}_{\perp}) = \text{Re} \sqrt{2m_F \left[\mu_F - g_{BF}
      A \rho_B^0 - \epsilon (\ve{k}_\perp) \right] }\; .
\end{equation*}

Finally, the \emph{thermodynamic} grand-canonical free energy density
is obtained by finding the global minimum of the potential
$f(\mu_B^{1D}, \mu_F, \rho_B^0)$ with respect to the boson density
$\rho_B^0$:
\begin{equation}
  \Omega(\mu_B^{1D}, \mu_F) = \min_{\rho_B^0 = n_B} f(\mu_B^{1D},
  \mu_F, \rho^0_B)\; ,
\label{eq:minim}
\end{equation}
where $n_B$ denotes the \emph{equilibrium} lineal boson density. Note
that, since Bose-Einstein condensation is not allowed in 1D
interacting boson systems~\cite{RMP_Cazalilla}, the boson density
cannot be regarded as the condensate density, i.e., the square of the
condensate order parameter. In addition, the \emph{equilibrium} 1D
fermionic density can be evaluated from
\begin{equation}
  n_F = \Frac{1}{N} \sum_{\ve{k}_{\perp} \in BZ}
  \Frac{k_{Fx}(\ve{k}_{\perp})}{\pi}\; .
\end{equation}
As explained in the next section, we can thus now evaluate the system
equilibrium phase diagram either in chemical potential or in density
space.

\begin{figure}
\includegraphics[width=1.0\linewidth,angle=0]{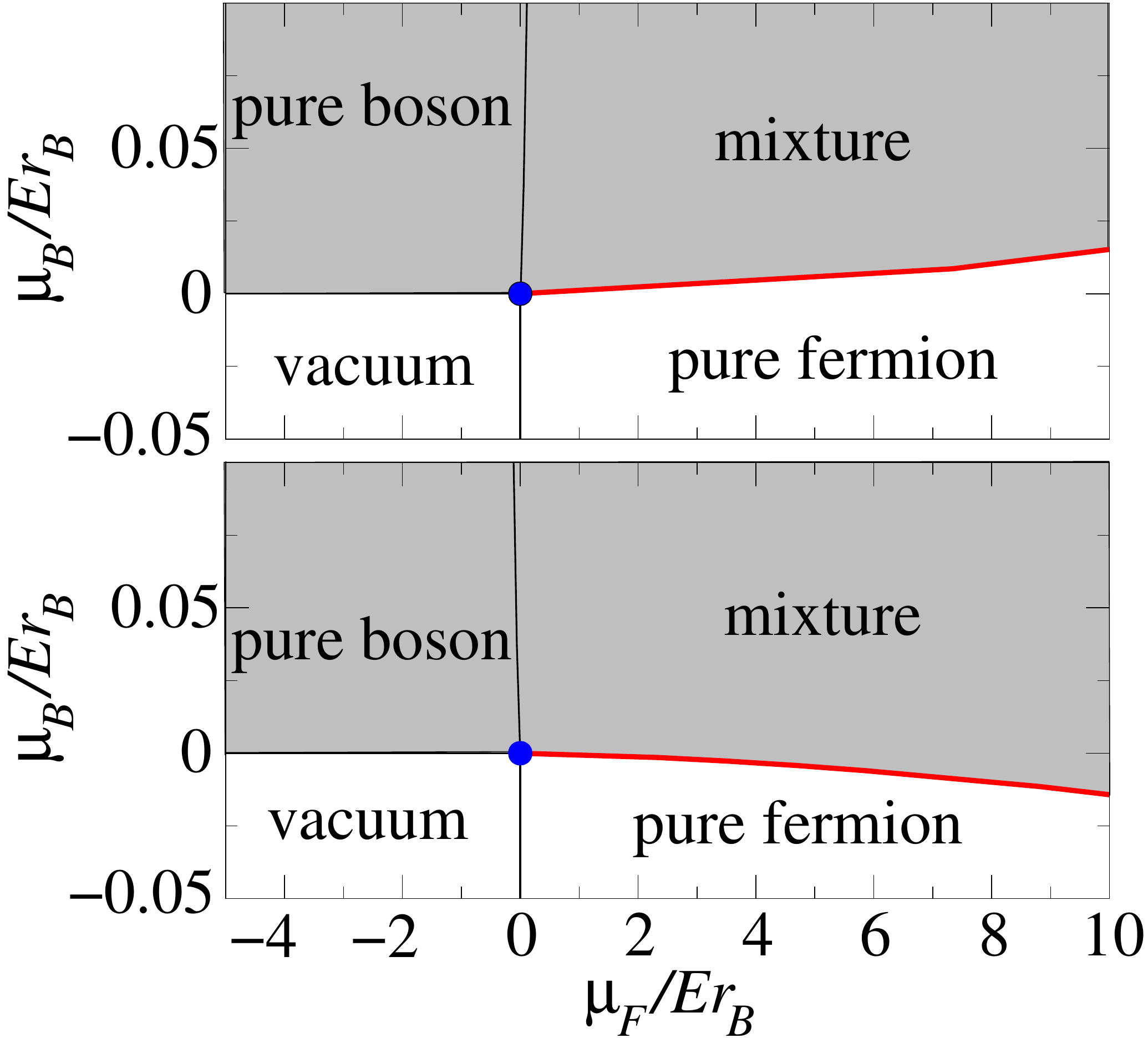}
\caption{(Color online) Zero temperature phase diagrams for a
  \emph{repulsive} ($a_{BF}>0$, top panel) and \emph{attractive}
  $a_{BF}<0$ (bottom panel) Bose-Fermi mixture with fixed
  dimensionless hopping amplitude $\tilde{t}=37.8$ and fixed
  interaction parameter $\zeta=0.23$ (the parameters of this
  calculation correspond to a $^{174}$Yb-$^6$Li system, see main
  text). The diagrams are plotted vs. the boson $\mu_B^{1D}/Er_B$ and
  fermion $\mu_F/Er_B$ chemical potentials, where $Er_B$ is the boson
  recoil energy --- note that $\mu^{1D}_B/Er_B \ll 1$, thus ensuring
  that the bosons remain confined to 1D throughout.  The thin solid
  black lines correspond to 2nd order (i.e. continuous) phase
  transitions between either the vacuum and the pure boson/fermion
  phases or between the pure boson and the homogeneous mixed
  phase. The thick solid (red) line corresponds to a first order
  transition between the pure fermion and the mixed phase. Second and
  first order lines meet at at the tricritical point at $\mu_F = \mu_B
  = 0$ (filled blue circle). The region with finite boson density is
  the light gray shaded region. In both cases phase separation can
  only occur between a mixed and pure fermion phases.}
\label{fig:cheft}
\end{figure}
%
\section{Phase diagram}
\label{sec:phase}
In this section, we obtain the phase diagram of the
mixed-dimensionality system by minimizing the free energy introduced
above in Eq.~\eqref{eq:freen} with respect to the boson density
$\rho_B^0$ for fixed $\mu_B^{1D}$ and $\mu_F$.  The free energy
$f(\mu_B^{1D}, \mu_F, \rho^0_B)$ also depends on several other
parameters, such as $g_{BF}$, $g_{BB}^{1D}$, and $t$, as well as the
particle masses $m_{B},m_{F}$. Thus, it is convenient to simplify the
description of the system by considering the following minimal set of
four independent dimensionless parameters:
\begin{align}
  \tilde{\mu}_B^{1D} &= \frac{2 m_B \mu_B^{1D}}{c^2} &
  \tilde{\mu}_F &=\frac{2 m_F^{1/3} m_B^{2/3} \mu_F}{c^2}\; \\
  \tilde{t}&=\frac{2 m_F^{1/3} m_B^{2/3} t}{c^2} &
  \zeta &= \frac{c}{2 m_B g_{BB}^{1D}}\; ,
\end{align}
where $c=2 A m_F^{1/3} m_B^{2/3} |g_{BF}|$. Thus, 
the dimensionless interaction parameter $\gamma$ can we rewritten as
$\gamma = (2\zeta \tilde{\rho}_B^0)^{-1}$, with the dimensionless
boson density given by $\tilde{\rho}_B^0 =\rho^0_B/c$.  In terms of
the above dimensionless quantities, the free energy  takes the form:
\begin{equation}
  \tilde{f} = (\tilde{\rho}_B^0)^3 e(\gamma) - \tilde{\mu}_B^{1D}
  \tilde{\rho}_B^0 - \Frac{1}{N} \sum_{\ve{k}_{\perp} \in BZ}
  \Frac{\tilde{k}_{Fx}^3 (\ve{k}_{\perp})}{3\pi m_F} \; ,
\label{eq:free2}
\end{equation}
where $\tilde{k}_{Fx} (\ve{k}_{\perp}) = \text{Re} \sqrt{\tilde{\mu}_F
  - \text{sign}(g_{BF}) \tilde{\rho}_B^0 - \tilde{\epsilon}
  (\ve{k}_\perp)}$ and $\tilde{\epsilon} (\ve{k}_\perp) = 2\tilde{t}
[2 - \cos(k_{y}d) -\cos(k_{z}d) ]$.
 
\subsection{Mixed dimensions}
\label{sec:mixdi}
We explicitly consider here the case of mixed dimensions, while later
in Sec.~\ref{sec:1Dlim}, we will derive the phase diagram for the case
of pure 1D. We have numerically minimized $\tilde{f}$ by fixing the
value of the dimensionless interaction parameter, $\zeta$, and hopping
amplitude, $\tilde{t}$.  In order to make contact with ongoing as well
as future experiments~\cite{Hansen:2013gp, 2011PhRvL.106o3201I,
  2010ApPhB..98..691O}, we consider the specific case of a Bose-Fermi
system consisting of a light fermionic species such as $^6$Li and a
heavy bosonic species like $^{174}$Yb.  When this system is loaded in
a sufficiently deep 2D optical lattice, the large boson to fermion
mass ratio ($m_B/m_F \simeq 29$) is enough to suppress the hopping
between tubes of bosons, while allowing fermions to hop between the
tubes. This makes it possible to realize our initial assumption of
mixed dimensionality for the system. Indeed, in the limit of a deep
lattice, the fermion hopping amplitude in the tight-binding
approximation of Eq.~\eqref{eq:fermh} can be expressed in terms of the
optical potential strength $V^{F}_0$ and the Fermi recoil energy
$Er_F$, where $Er_{\alpha}=2\pi^2/(m_{\alpha} \lambda^2)$ and $\lambda
= 2d$ is the wavelength of the laser generating the optical lattice
potential~\cite{Bloch:2008gl}:
\begin{equation}
  t\simeq \frac{4 Er_{F}}{\sqrt{\pi}}
  \left(\Frac{V_0^F}{Er_F}\right)^{3/4} e^{-2\sqrt{V_0^F/Er_F}}\; .
\end{equation}
For a laser wavelength $\lambda=1064$~nm, the deep lattice condition
is achieved by making $V^B_0 \simeq 40 Er_B$ (for this system $V^F_0
\simeq 2 V^B_0$~\cite{Hansen:2013gp,private_taka}).  Furthermore, the
boson-boson scattering ~\eqref{eq:scatt} length has been
experimentally estimated to be
$a_{BB}=104.9~a_0$~\cite{Kitagawa:2008eh} (where $a_0$ is the Bohr
radius). The 1D interaction strength $g_{BB}^{1D}$ can be obtained
from~\cite{PhysRevLett.81.938}:
\begin{equation}
  \Frac{1}{g_{BB}^{1D}} = \frac{m_B \ell_B}{2} \left(
  \frac{\ell_B}{a_{BB}} - C\right) \; ,
\label{eq:olsha}
\end{equation}
where $C \simeq 1.0326$ and $\ell_B = 67.34$~nm~\footnote{Note that,
  approximating the confining optical lattice as harmonic, one has
  that the trapping frequencies for bosons and fermions are given by
  $\omega_{\alpha}=\ell_{\alpha}^2/(m_{\alpha})=2Er_{\alpha}
  \sqrt{V_0^{\alpha} / Er_{\alpha}}$}. For these parameters,
$\ell_F=131.40$~nm, and therefore, using Eq.~\eqref{eq:acon2},
$A=1.46017\times 10^{13}~\mathrm{m^{-2}}$.  Finally, we set
$|a_{BF}|=13~a_0$~\cite{Hansen:2013gp}, and allow for both positive
and negative signs for $a_{BF} $, i.e.  for the Bose-Fermi
interactions to be repulsive or attractive. Using these values, the
dimensionless interaction and hoping parameters are $\tilde{t}=37.8$
and $\zeta = 0.23$.


The phase diagrams for repulsive ($a_{BF} > 0$) and attractive
($a_{BF} < 0$) Bose-Fermi interactions are shown in
Fig.~\ref{fig:cheft} as a function of the boson and fermion chemical
potentials.  In both cases, a qualitatively similar structure emerges:
The transition between the pure fermion and the phase where boson and
fermions form a homogeneous mixture (mixed phase) is first order
(thick solid red line). For the particular values of parameters chosen
in Fig.~\ref{fig:cheft} to describe the $^{174}$Yb-$^6$Li mixture, we
numerically find that the transition is weakly first order close to
the origin $(\mu_F,\mu_B^{1D})=(0,0)$, i.e., the chemical potential
values for which the slope of the free energy at $\rho^0_B = 0$
changes sign are very close to the values of chemical potentials at
the transition. On the other hand, the transition between the pure
boson and the mixed phases (thin solid black line) is second order,
i.e. continuous. This transition coincides with the locus of points
where the system first develops a Fermi surface, i.e., $\tilde{\mu}_F
- \text{sign}(g_{BF}) \tilde{\rho}_B^0 = 0$, and therefore $\mu_F >0$
($\mu_F <0$) for repulsive (attractive) interactions. In
App.~\ref{app:fermi}, we carry on an expansion for small fermion
density which allows to establish the nature of the phase transitions
where the number of Fermi surfaces changes from zero to one. There, we
argue that this transition is continuous because of the scaling that
the Fermi kinetic energy has with the fermion density in 3D, while it
would be first order if the fermion would also move in strictly 1D
like the bosons. This result is also in agreement with the conclusion
reached by Viverit \emph{et al.}  for Bose-Fermi mixtures in
3D~\cite{Viverit:2002hg}, where they show that phase separation
between a mixed phase and a pure boson phase cannot be realized in 3D.

Finally, the transitions between the vacuum, corresponding to zero
density of both fermions and bosons, and either the pure boson or
fermion phases (thin solid black lines in Fig.~\ref{fig:cheft}) are
continuous, as they correspond to the filling of a
band~\cite{qpt-sachdev}. Therefore, the first order line separating
the pure fermion and mixed phases terminates at the origin
$(\mu_F,\mu_B^{1D})=(0,0)$ in a tricritical point (filled blue
circle), where the first order transition becomes second order.

\begin{figure}
\includegraphics[width=1.0\linewidth,angle=0]{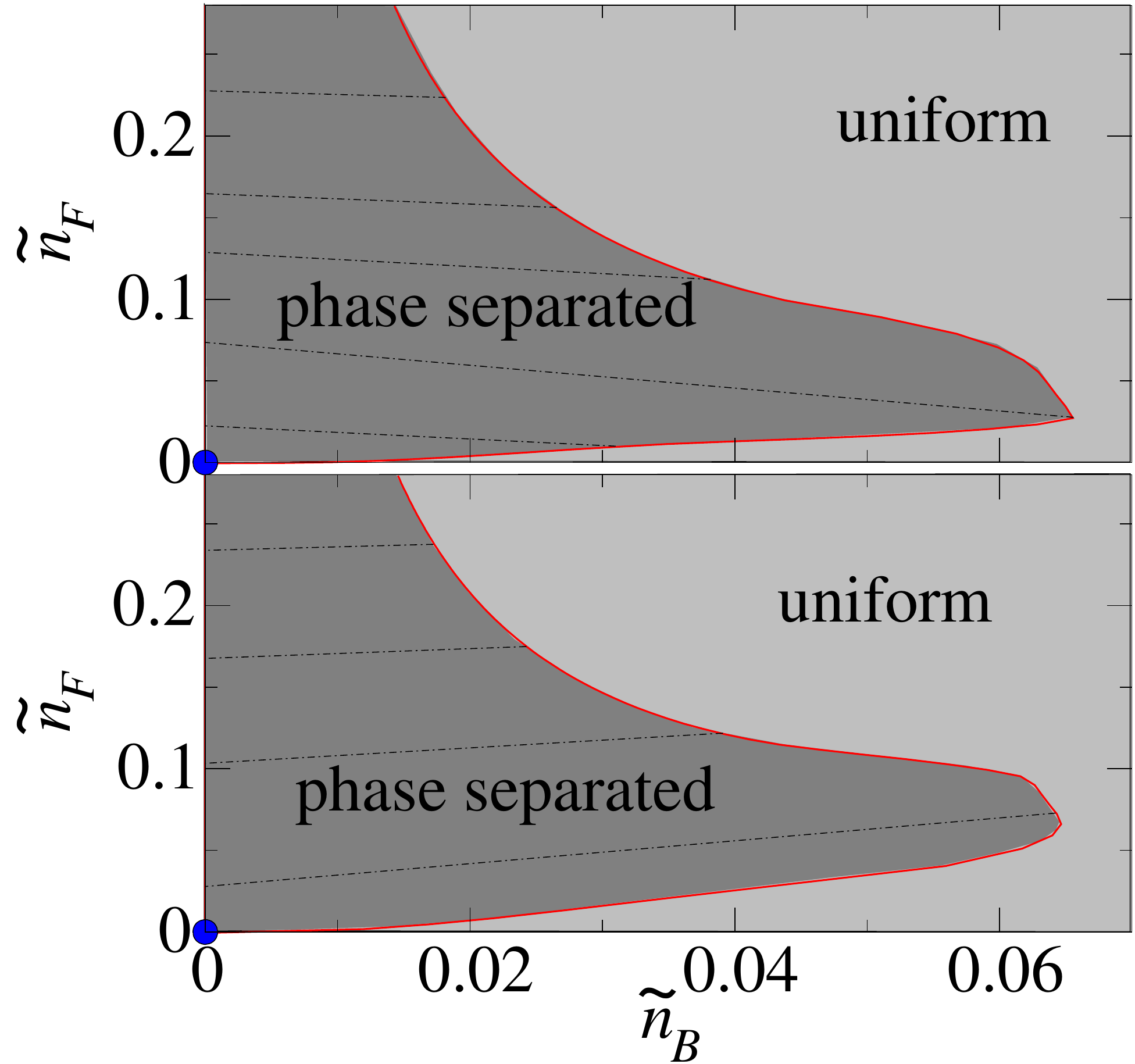}
\caption{(Color online) Phase diagrams in the density plane
  $(\tilde{n}_B, \tilde{n}_F)$. In order to obtain a more accurate
  estimate of the phase boundaries, we have used the parameters
  $\tilde{t} = 0.023$ and $\zeta = 0.27$. The corresponding phase
  diagrams in the chemical potential plane (not shown for brevity)
  display the same features and phase topology as the diagrams shown
  Fig.~\ref{fig:cheft}, which are computed for the experimentally
  relevant a $^{174}$Yb-$^6$Li system, but for which the phase
  boundary in the density plane proved much harder to determine
  numerically.  The top panel corresponds to a repulsive Bose-Fermi
  interactions, whereas in the bottom corresponds to attractive
  interactions. In both cases, the system can either be in a uniform
  mixed phase (lightly gray shaded region) or, by undergoing a first
  order transition, it can be in a phase separated state (dark gray
  shaded region), where a pure fermion and mixed phase coexist. The
  dot-dashed lines connect points on the first order boundary with the
  same values of the chemical potentials.}
\label{fig:denft}
\end{figure}

A first order transition in the phase diagram in chemical potential
space implies that the system exhibits phase separation in density
space, where, rather than fixing the chemical potentials $\mu_F$ and
$\mu_B^{1D}$, one fixes the boson $n_B$ and fermion $n_F$ densities
(see Fig.~\ref{fig:denft}). We obtain therefore that, for finite
inter-tube hopping $t$, phase separation is only possible between pure
fermion and mixed phases (dot-dashed black lines in
Fig.~\ref{fig:denft}). In Sec.~\ref{sec:1Dlim}, we will see that this
fact is related to the dimensionality where fermions move and that the
situation drastically changes for strictly 1D, e.g., when the hopping
$t$ for fermions is reduced to zero. Note also that, for attractive
interactions, in contrast to a 3D Bose-Fermi mixtures in the absence
of the lattice~\cite{PhysRevA.61.053605,pethicsmith}, the system is
found to exhibit phase separation rather than collapse. This result
was also obtained in Ref.~\cite{Marchetti:2009de}, by treating the
boson-boson interactions within the mean-field approximation. However,
different from that work, the non-perturbative treatment of the boson
interactions employed here, yields a first order transition between
the pure fermion and mixed phases.

The absence of a continuous phase transition between the pure fermion
and any phase with a finite density of bosons can be qualitatively
understood by making an analogy with the Landau theory of phase
transitions and considering the series expansion of the free energy
$\tilde{f}$ in~\eqref{eq:free2} for small values of the boson density
$\tilde{\rho}_B^0$. In this limit, for fixed $\zeta^{-1}\propto
g^{1D}_{BB}$, we have that $\gamma=(2\zeta \tilde{\rho}_B^0)^{-1} \gg
1$, and therefore the Bose gas is essentially fermionized and close to
a Tonks gas.  Thus, we can use the following asymptotic formula for
the boson energy~\cite{Lieb:1963vu,Cazalilla:2003io}:
\begin{equation*}
  e(\gamma) \Simiq_{\gamma \to \infty} e_{TG} \left[1 -
    \frac{4}{\gamma} + O(\gamma^{-2})\right]\; ,
\end{equation*}
where $e_{TG} =  \pi^2/3$. Note that this expression implies that 
the boson contribution to the free energy 
grows as $(\rho^0_B)^3$. This yields the following series expansion for
the free energy at small boson density:
\begin{equation}
  \tilde{f} = \tilde{f}_0 + \tilde{f}_2 \tilde{\rho}_B^0 + \tilde{f}_2
  (\tilde{\rho}_B^0)^2 + \tilde{f}_6 (\tilde{\rho}_B^0)^3 + \cdots \;
  ,
\label{eq:expan}
\end{equation}
where the coefficients of the expansion are given by:
\begin{align*}
  \tilde{f}_2 &= \text{sign}(g_{BF}) C_1 - \tilde{\mu}_B^{1D}\; ,\\
  \tilde{f}_4 &= -C_{-1}/4\; ,\\
  \tilde{f}_6 &= e_{TG} - \text{sign} (g_{BF}) C_{-3} /24\; ,
\end{align*}
and where $C_{n} = \frac{1}{\pi N} \sum_{\ve{k}_{\perp} \in BZ}
\text{Re} \left[\sqrt{\tilde{\mu}_F - \tilde{\epsilon}
    (\ve{k}_\perp)}\right]^{n} \ge 0$ ($|n|$ is odd). In addition,
note that $C_{n} = 0$ if $\mu_F = 0$. Thus, for $\mu_F > 0$, a
continuous phase transition cannot take place because the coefficient
$\tilde{f}_4$ of the above expansion is always negative meaning that
for $\tilde{f}_2 > 0$ the free energy must eventually decrease away
from the origin where $\tilde{\rho}_B^0 = 0$ before it can rise again
($\tilde{f}_6 > 0$ is assumed, for stability). Thus, the free energy
develops a local minimum for $\tilde{\rho}_B^0 \neq 0$, which
eventually can be tuned to be degenerate with the local minimum at
$\rho^0_B = 0$.  It is worth comparing this situation with the result
of the mean-field treatment of bosons interactions carried out by one
of us in Ref.~\cite{Marchetti:2009de}, where it was found that
$\tilde{f}_4 = (2 - \zeta C_{-1})/(4\zeta)$, thus allowing both second
and first order phase transitions to occur for $\mu_F>0$ by tuning
$\tilde{f}_2 = 0$ and $\tilde{f}_4 = 0$.

In addition, we can use the above expressions to understand the
emergence of a tricritical point, which corresponds to the conditions
$\tilde{f}_2 = 0 = \tilde{f}_4$ while $\tilde{f}_6>0$ for
stability. Since $C_n =0$ only for $\mu_F = 0$, we can conclude that a
stable tricritical point can only exist at the origin of the chemical
potential plane, i.e. for $(\mu_F,\mu_B^{1D})=(0,0)$. Close to the
tricritical point, the shape of the first order line can also obtained
analytically using the conditions $\tilde{f}_2>0$ and $\tilde{f}_4 =
-2 \sqrt{\tilde{f}_2 \tilde{f}_6}$. Note that for the choice of
parameters done to describe the $^{174}$Yb-$^6$Li mixture, we find
that, the true first order transition obtained numerically stays very
close to the one found analytically here. In addition, the transition
is weakly first order because of the large values of the coefficient
$\tilde{f}_6$ for which $\tilde{f}_2 \to 0$.

\begin{figure}
\includegraphics[width=1.0\linewidth,angle=0]{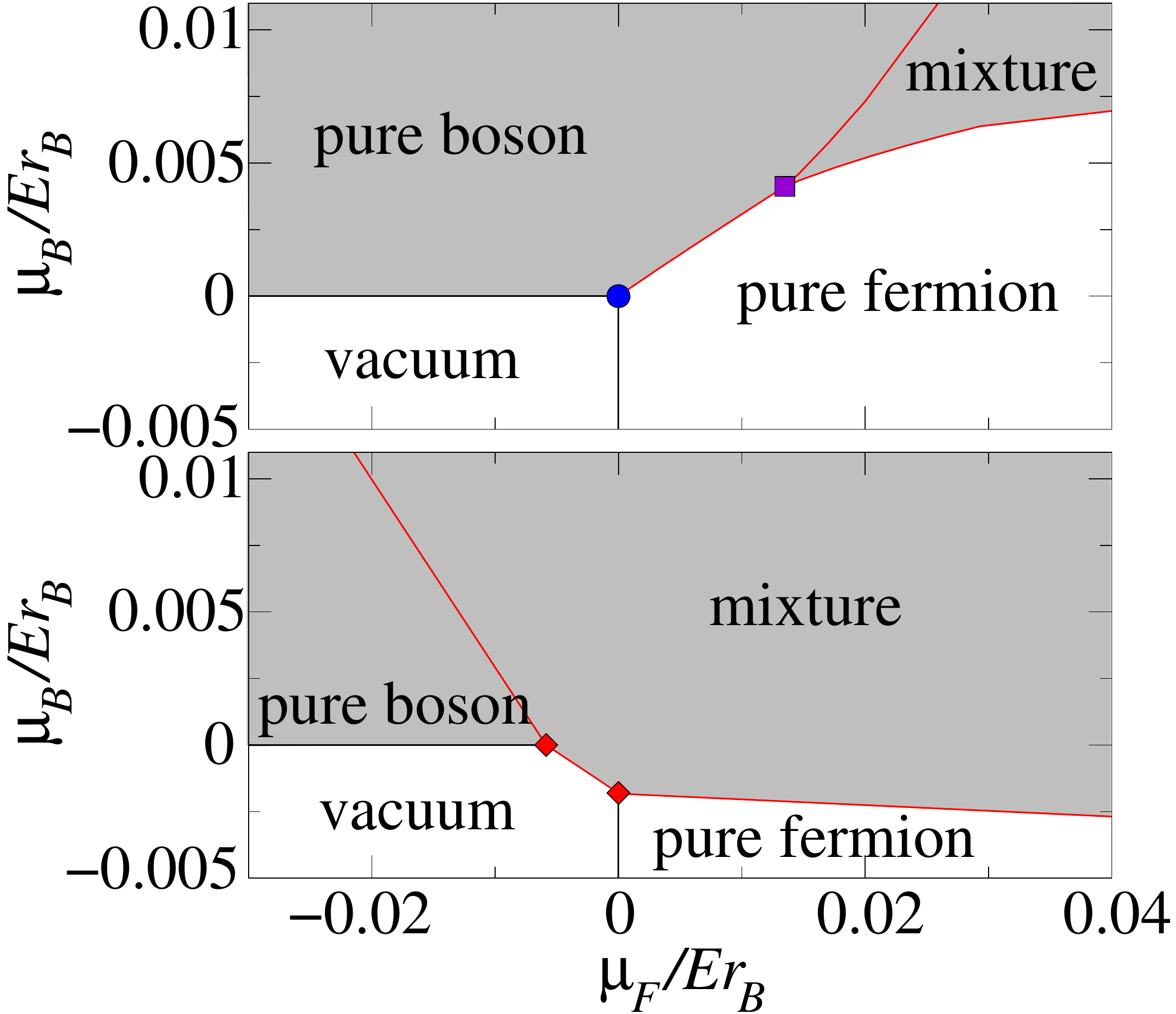}
\caption{(Color online) Zero temperature phase diagrams for a
  \emph{repulsive} ($a_{BF}>0$, top panel) and \emph{attractive}
  ($a_{BF}<0$, bottom panel) Bose-Fermi $^{174}$Yb-$^6$Li mixture in
  1D, i.e., for zero hopping amplitude $t=0$, and for $zeta=0.23$.
  Differently from the finite $t$ case (cf. Fig.~\ref{fig:cheft}), the
  transition between the pure boson and mixed phase becomes first
  order (thick red solid line), meeting with the other two first order
  transition lines (one between the pure fermion and mixed phase and
  the other between the pure fermion and the pure boson phases) at a
  triple point (filled violet square), where the three phases
  coexists. Bottom panel: For an attractive mixture there is neither a
  triple point nor a tricritical point, rather the first order line
  crosses the second order line at two critical end-points (filled
  [red] diamond) delimiting the region of phase separation between the
  vacuum and a mixed phase.}
\label{fig:phas3}
\end{figure}
%
\subsection{Pure 1D limit}
\label{sec:1Dlim}
Next, we focus on the pure 1D limit, i.e., the limit where the
fermions, like the bosons, cannot hop between the tubes
(i.e. $t=0$). The phase diagram in chemical potential resulting from
minimizing the free energy is shown in Fig.~\ref{fig:phas3}, for both
repulsive and attractive Bose-Fermi interactions. It can be seen that,
in the pure 1D limit, for both repulsive and attractive interspecies
interactions, all transitions (except for the trivial ones from the
vacuum phase) are discontinuous. In particular, the transition between
the pure boson to the mixed phases, which was found to be continuous
in the mixed dimensional system, becomes discontinuous as soon as the
fermions are confined to 1D. In App.~\ref{app:fermi} we carry on an
expansion for small fermion density that allows us to establish the
nature of the phase transitions where the number of Fermi surfaces
changes from zero to one. As shown there, the main difference between
the mixed-dimensional case illustrated in the previous section and the
pure 1D limit analyzed here can be traced down to the different
scaling of the Fermi kinetic energy with the lineal fermion density
$\rho^0_F$ in 1D and 3D. In particular, whereas in 1D the Fermi
kinetic energy scales as $(\rho^0_F)^3$, in 3D it scales more slowly
as $(\rho^{0}_F)^{5/3}$.

Furthermore, similarly to what was found in the previous section, the
fermionization of bosons in 1D also renders the transition between the
pure fermion and mixed phases discontinuous. These results are
compatible with the previous results obtained by Das in
Ref.~\cite{Das:2003jr}, where was found that the transition between
the pure boson and fermion phases is discontinuous thus leading to
phase separation.

\begin{figure}
\includegraphics[width=1.0\linewidth,angle=0]{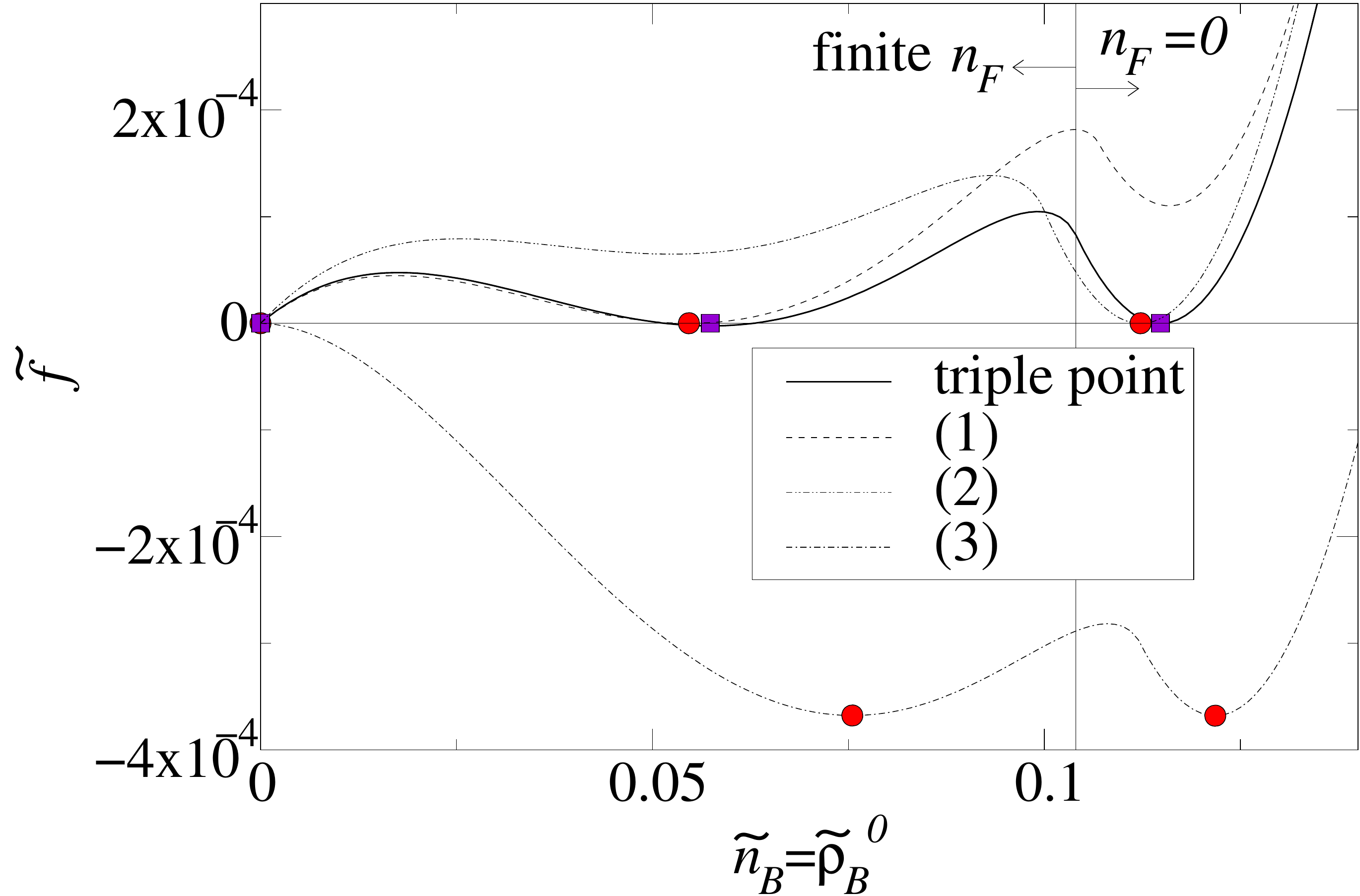}
\caption{(Color online) Dimensionless free energy density $\tilde{f}$
  plotted versus the dimensionless bosonic density $\tilde{n}_B =
  \tilde{\rho}_B^0$ for fixed dimensionless hopping strength
  $\tilde{t} = 0.023$, fixed interaction parameter $\zeta = 0.27$, and
  a 1D repulsive Bose-Fermi system, corresponding to the density phase
  diagram shown in the top panel of Fig.~\ref{fig:denzt}. The values
  of the chemical potentials are fixed at the three different first
  order transition lines near the triple point $(\mu_B^{1D}/Er_B,
  \mu_F/Er_B)\simeq(35.3, 11.2)$, which describes the state where
  phase separation occurs between the three phases, pure boson, pure
  fermion and mixed phase: $(1)\simeq(36.3, 11.4)$ describe phase
  separation between pure fermion and mixed phase; $(2)\simeq (33.9,
  10.8)$ phase separation between pure fermion and pure boson; finally
  $(3)\simeq (38.1, 12.4)$ phase separation between mixed and pure
  boson.}
\label{fig:freen}
\end{figure}

\begin{figure}
\includegraphics[width=1.0\linewidth,angle=0]{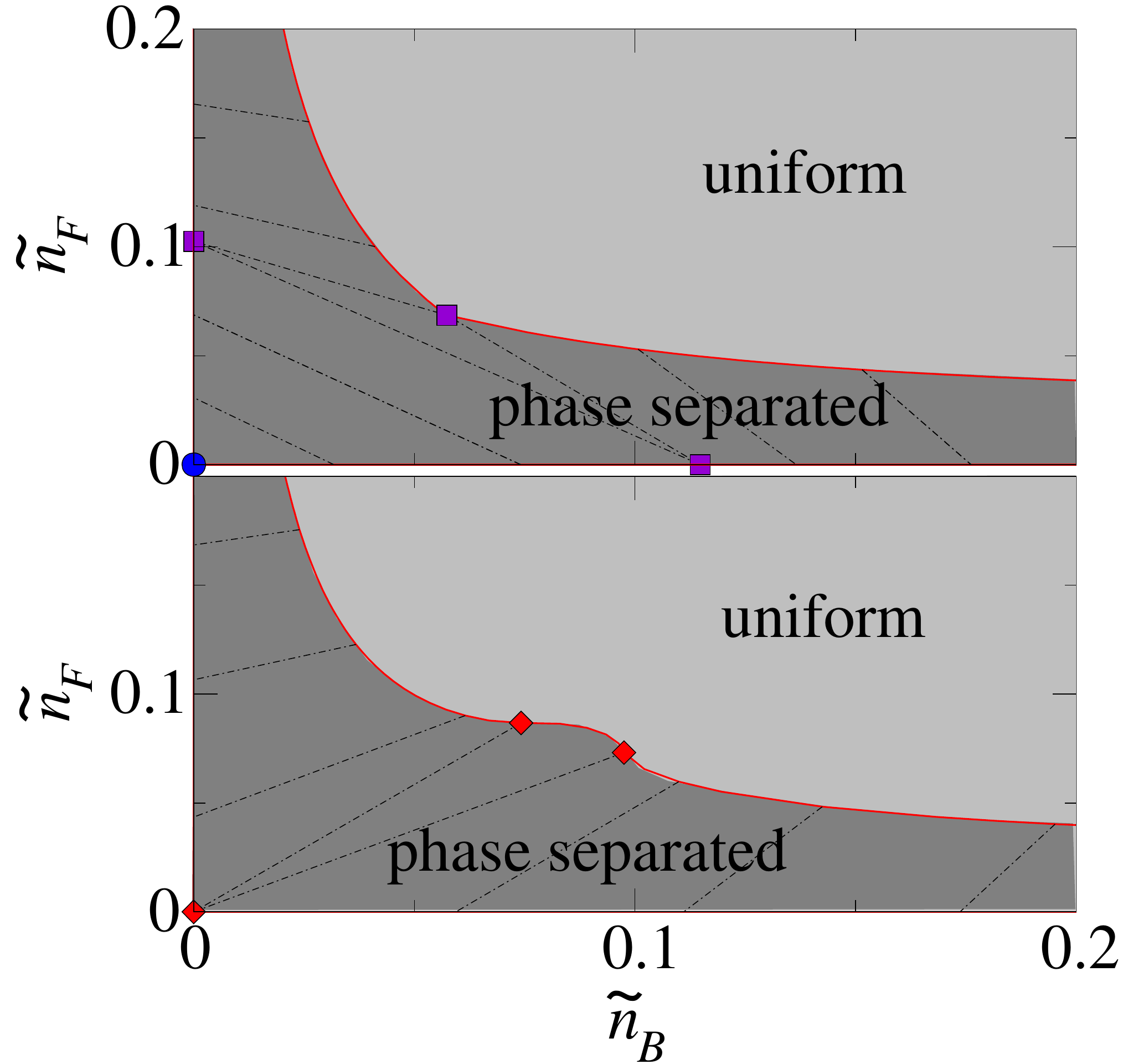}
\caption{(Color online) Phase diagrams in the density plane
  $(\tilde{n}_B, \tilde{n}_F)$ for $\zeta = 0.27$ and zero hopping
  strength $t = 0$ (same remarks as for Fig.~\ref{fig:denft} apply).
  Top panel: for repulsive interactions in the 1D limit, phase
  separation occurs, not only between a pure fermion and a mixed
  phase, but also between a pure boson and pure fermion phase, as well
  as between a pure boson and a mixed phase. Bottom panel: In
  contrast, for the attractive case, there is no phase separation
  between pure phases, rather, it exists a region of phase separation
  between the vacuum and a mixed phase.}
\label{fig:denzt}
\end{figure}

The main difference between the repulsive and the attractive cases is
the way the transitions across which the density of fermions changes
connect with the transition between the pure fermion and mixed phase.
In particular, in the repulsive case (upper panel of
Fig.~\ref{fig:phas3}), the three first order transition lines (thick
red solid curves) between pure boson-pure fermion, pure fermion-mixed,
and pure boson-mixed phases meet at a triple point (filled violet
square symbol). At this triple point, the three phases coexist since
the free energy exhibits three degenerate local minima (see
Fig.~\ref{fig:freen}). On the other hand, for the attractive case (see
lower panel of Fig.~\ref{fig:phas3}), the triple point is
absent. Instead, two critical end points (filled red diamonds) appear.
In the density phase diagram, the critical end points delimit a
triangularly-shaped region (see bottom panel of Fig.~\ref{fig:denzt}),
where phase separation occurs between the vacuum and mixed
phases. Similarly to conclusion reached in
Ref.~\cite{Marchetti:2009de}, this region can be regarded as a remnant
of the collapse that occurs in the absence of a lattice in 3D
Bose-Fermi mixtures with sufficiently large attractive
interactions~\cite{PhysRevA.61.053605,pethicsmith}.

Let us finally remark that 1D Bose-Fermi mixtures in the exactly
solvable limit of equal masses (i.e. $m_B = m_F$) and equal
interactions strengths ($g_{BB}^{1D} = g_{BF})$ have been analyzed in
Ref.~\cite{Imambekov:2006ib}. By relying on a linear stability
analysis, which requires that the compressibility matrix must be
positively defined for any strength of the interactions, the authors
of Ref.~\cite{Imambekov:2006ib} concluded that the system is always
stable against demixing. However, in this work, by \emph{globally
  minimizing} the system free energy, i.e., by looking for the global
energy minima, we find that phase separation occurs in a rather broad
region of the density phase diagram and, in particular, it always
occurs for small bosons and fermion densities in the pure 1D limit.

%
\section{Conclusions and Outlook}
\label{sec:concl}
To summarize, we have studied the equilibrium phase diagram of a
Bose-Fermi system in a mixed-dimensional geometry of an array of
coupled 1D tubes, where bosons are strongly localized along the 1D
tubes, while fermions can hop between the tubes. Because we treat the
boson interaction term non-perturbatively using the Bethe ansatz, we
have found that the transition between the pure Fermi phase and the
mixed phase is always first order. This implies that, for this system,
phase separation can take place between a pure fermion and a mixed
phase. However, the phase transition between pure boson and mixed
phases is found to be continuous for finite hoping amplitude of the
fermions and phase separation cannot take place between these two
phases. This contrasts with the results obtained by assuming that the
bosons form a quasi-condensate and thus applying a standard mean-field
treatment where the boson density is replaced with its expectation
value~\cite{Marchetti:2009de}. In that case, the free energy the
transition between the pure fermion and a mixed phases can be
continuous as well as discontinuous, depending on the system
parameters.

In the pure 1D limit, we found that the all transitions are first
order, except for the trivial ones between the vacuum state and the
pure fermion or boson phases. The main differences introduced by the
sign of the interaction are the existence of a triple point for
repulsive interactions, and the appearance of two critical end points
delimiting a first order line between the mixed and the vacuum phases.

Thus, the main difference between the 1D limit and the mixed
dimensional system is the change in the character of the transition
between the pure boson and mixed phases. We have argued (see
App.~\ref{app:fermi}) that this difference is a consequence of the
different scaling of the Fermi gas kinetic energy contribution to the
free energy function with the fermion density.

Before commenting on the relevance of our findings for the
experiments using $^{174}$Yb-$^{6}$Li mixtures, it is interesting to note
that, because we have applied the Bethe ansatz to the 1D Bose
Hamiltonian $\hat{H}_{B}^{mf}$~\eqref{eq:mfbos}, we can extend our
analysis to to Fermi-Fermi system consisting of a light and a heavy
atom. In fact, in the limit $g_{BB} \to +\infty$ where $\gamma \to
\infty$, the boson energy, $e(\gamma)$ becomes identical to that of a
free Fermi gas~\cite{girardeau}. By taking this limit of the
Bethe-energy in Eq.~\eqref{eq:free2}, we found that the phase diagrams
for both repulsive and attractive Fermi-Fermi interactions evaluated
in the same way as before are qualitatively similar to those displayed
in Fig.~\ref{fig:cheft} for the Bose-Fermi mixture and that the
differences are small and only quantitative~\footnote{In particular,
  we have checked that for fixed interaction parameter $\zeta=0.23$,
  the relative difference between the exact Bethe-ansatz energy
  $e(\gamma)$ and the Tonks gas limit $e_{TG}$ can be only roughly up
  to $12\%$ for typical densities $\tilde{\rho}_B^0$ involved in the
  first order transition. As a result, we obtain that the phase
  diagrams for both repulsive and attractive interactions maintain the
  same qualitative features.}.

The actual experimental systems are rendered inhomogeneous by the
existence of harmonic trapping. For mixed dimensionality systems, in
general, and the $^{174}$Yb-$^{6}$Li system~\cite{Hansen:2013gp,
  2011PhRvL.106o3201I,2010ApPhB..98..691O} loaded in an anisotropic
optical lattice in particular, we can rely on the local density
approximation and deduce the main implications for experiments from
the phase diagrams in the chemical potential space shown in
Fig.~\ref{fig:cheft} and in Fig.~\ref{fig:phas3}. For a given trap
frequency and total atom numbers $N_F$ and $N_B$, it is possible to
determine the range of values of the chemical potentials $\mu_F$ and
$\mu_B^{1D}$ of the phase diagrams in Figs.~\ref{fig:cheft} and
\ref{fig:phas3} being sampled by the trapped system. This range
determines a region in the chemical phase diagram that contains the
possible phases that will coexist in the trap.

Lastly, we comment on the accuracy of the above mean-field approach
that we have employed for the Bose-Fermi interaction term. Indeed, as
any other mean-field theory, it neglects fluctuations. We expect
fluctuations to be especially important close to a phase transition.
Nevertheless, as shown later in Sec.~\ref{sec:phase}, our calculations
indicate that many of the phase transitions in the mixed
dimensionality Bose-Fermi system (cf.  \ref{sec:mixdi}) and in the
pure 1D limit are discontinuous. This means that, even very close to
the transition point, fluctuations of the boson and fermion densities
are typically suppressed and thus it can be expected that the
mean-field theory to give a reliable picture. However, we also
numerically observed that some of the transitions are only weakly
first order.  In other cases, however, the transition was found to be
continuous.  Therefore a careful assessment of the effect of
fluctuations will be required but will not be pursued here.

Furthermore, within the above mean-field approach, we also have
neglected the inter-tube couplings as well as the effect of the bosons
on the fermion properties, which are as a non-interacting Fermi
gas. These are also concerns that deserve to be investigated in the
future work. Here, we have assumed that such effects are relatively
weak and can only become important only for large values $|a_{BF}|$
and/or very low temperatures that may not be easily achievable under
current experimental conditions.

Beyond the assessment the accuracy of the present mean-field approach,
another interesting direction is to apply the methods developed here
to mixed dimensionality systems where the 1D bosons interact via
longer range interactions, such as dipolar gases or are tuned to the
so-called the super-Tonks
regime~\cite{Batchelor:2005gt,Chen:2010bn,Haller:2009jr,Qi:2013jd,Valiente:2012co,RMP_Cazalilla}.

\acknowledgments We are grateful to M. Parish and Y. Takahashi for
stimulating discussions and suggestions. FMM acknowledges financial
support from the programs Ram\'on y Cajal, the Spanish MINECO
(MAT2011- 22997), CAM (S-2009/ESP-1503), and Intelbiomat (ESF). EM
acknowledges support from the CSIC JAE-predoc program, co-financed by
the European Science Foundation. EM and MAC also acknowledge the
support of the Basque Departamento de Educaci\'on, the UPV/EHU (Grant
No.  IT-366-07), and the Spanish MINECO (Grant
No. FIS2010-19609-CO2-02).

\appendix
\section{Mean-field approximation}
\label{app:avera}
In this appendix, we provide the intermediate steps necessary to
obtain the mean-field Hamiltonian, Eq.~\eqref{eq:meanf}.  First of
all, let us evaluate the expectation value of the fermion density
operator in a tube $\hat{\rho}_{F\ve{R}}(x)$. By substituting the
expression~\eqref{eq:fexpa} into~\eqref{eq:fsing} and using that, for
a non-interacting Fermi gas, $\langle \hat{f}^{\dag}_{\ve{k}}
\hat{f}^{\phantom{\dag}}_{\ve{k}'} \rangle = \delta_{\ve{k},\ve{k}'}
n_{\ve{k}}$, with $n_{\ve{k}}$ being the Fermi-Dirac distribution
function, we obtain:
\begin{equation}
  \langle \hat{\rho}_{F\ve{R}}(x) \rangle = \Frac{1}{L} \sum_{\ve{k}}
  n_{\ve{k}} \int d\ve{r}_{\perp} |\varphi^{\phantom{*}}_{\ve{R}}
  (\ve{r}_{\perp})|^2 |\phi_{\ve{k}_{\perp}} (\ve{r}_{\perp})|^2 \; .
\label{eq:faver}
\end{equation}
Further, we can assume that the fermions occupy only the lowest Bloch
band of the square lattice, thus the Bloch wave function
$\phi_{\ve{k}_{\perp}} (\ve{r}_{\perp})$ can be developed in terms of
Wannier functions $w^{\phantom{*}}_{\ve{R}} (\ve{r}_{\perp})$ localized
at the $\ve{R}$-th tube:
\begin{equation}
  \phi_{\ve{k}_{\perp}} (\ve{r}_{\perp}) = \Frac{1}{\sqrt{N}}
  \sum_{\ve{R}} e^{i \ve{k}_\perp \cdot \ve{R}}
  w^{\phantom{*}}_{\ve{R}} (\ve{r}_{\perp})\; ,
\end{equation}
where $N=N_y N_z$. Therefore, Eq.~\eqref{eq:faver} reads:
\begin{multline}
  \langle \hat{\rho}_{F\ve{R}}(x) \rangle = \Frac{1}{L} \sum_{\ve{k}}
  n_{\ve{k}} \frac{1}{N} \sum_{\ve{R}',\ve{R}''} \int
  d\ve{r}_{\perp} e^{i \ve{k}_\perp \cdot (\ve{R}'-\ve{R}'')}\\
  \times w_{\ve{R}'}^* (\ve{r}_{\perp}) w^{\phantom{*}}_{\ve{R}''}
  (\ve{r}_{\perp}) |\varphi^{\phantom{*}}_{\ve{R}} (\ve{r}_{\perp})|^2
  \\
  \simeq \Frac{N_F}{L} \frac{1}{N} \int d\ve{r}_{\perp}
  |w^{\phantom{*}}_{\ve{R}} (\ve{r}_{\perp})|^2
  |\varphi^{\phantom{*}}_{\ve{R}} (\ve{r}_{\perp})|^2 = \rho^0_F A\; ,
\end{multline}
where $N_F = \sum_{\ve{k}} n_{\ve{k}}$, $\rho^0_F$ is the 1D density
of fermions~\eqref{eq:linde}, and
\begin{multline}
  A=N \int d\ve{r}_{\perp} |\varphi^{\phantom{*}}_{\ve{R}}
  (\ve{r}_{\perp})|^2 |\phi_{\ve{k}_{\perp}} (\ve{r}_{\perp})|^2\\
  \simeq \Frac{1}{N} \int d\ve{r}_{\perp} |w^{\phantom{*}}_{\ve{R}}
  (\ve{r}_{\perp})|^2 |\varphi^{\phantom{*}}_{\ve{R}}
  (\ve{r}_{\perp})|^2\; .
\label{eq:aexpr}
\end{multline}
In this derivation, we have used the fact that the boson Wannier
orbital $\varphi^{\phantom{*}}_{\ve{R}} (\ve{r}_{\perp})$ is strongly
localized around $\ve{r}_\perp = \ve{R}$, thus we have neglected the
contributions of the terms $\ve{R}'$ and $\ve{R}''$ different from
$\ve{R}$, because the corresponding wavefunction overlaps are
negligible. Note also that the constant $A$~\eqref{eq:aexpr} does not
depend on the tube index $\ve{R}$. Further, we can approximate
$\varphi^{\phantom{*}}_{\ve{R}}(\ve{r}_{\perp}) \simeq
e^{-|\ve{r}_{\perp}-\ve{R}|^2/2\ell^2_{B}}/(\sqrt{\pi} \ell_B)$ and
$w^{\phantom{*}}_{\ve{R}}(\ve{r}_{\perp}) \simeq
e^{-|\ve{r}_{\perp}-\ve{R}|^2/2\ell^2_{F}}/(\sqrt{\pi} \ell_F)$ with
$\ell_{B} < \ell_{F} \ll d$, therefore the constant $A$ is given by:
\begin{equation}
  A \simeq\frac{1}{\pi (\ell^2_{F} + \ell^2_{B})}\; .
\label{eq:acon2}
\end{equation} 

Taking into account that the mean-field averages of the boson and
fermion densities are given by the expressions~\eqref{eq:aver}, in
order to obtain the final expression of the Hamiltonian,
Eq.~\eqref{eq:meanf}, we need to deal with%
\begin{align}
 & \sum_{\ve{R}} \int dx \langle\hat{\rho}_{B\ve{R}}(x)\rangle
  \hat{\rho}_{F\ve{R}}(x)  \\
  &= \rho_{B}^0 \sum_{\ve{R}}\sum_{\ve{k}} \int d\ve{r}_{\perp}
  |\varphi^{\phantom{*}}_{\ve{R}} (\ve{r}_\perp)|^2
  |\phi_{\ve{k}_{\perp}} (\ve{r}_\perp)|^2 \hat{f}_{\ve{k}}^\dag
  \hat{f}^{\phantom{\dag}}_{\ve{k}} \\
  &\simeq \rho_{B}^0 A \sum_{\ve{k}}\hat{f}_{\ve{k}}^\dag
  \hat{f}^{\phantom{\dag}}_{\ve{k}}\; .
\end{align}
%

\section{Series expansion for small Fermi density}
\label{app:fermi}
In this appendix we want to carry on an expansion for small fermion
density so that to be able to establish the nature of the phase
transitions where the number of Fermi surfaces changes from zero to
one, such as the phase transition between a pure boson and a mixed
phase. This is done in the same spirit to the expansion for small
boson density conducted at the end of Sec.~\ref{sec:mixdi} that has
allowed us to establish that the transitions between pure fermion and
any phase with a finite density of bosons can only be first order.

The starting point is the mean-field Hamiltonian derived in
Eq.~\eqref{eq:meanf}, with the difference that now, to simplify the
analysis, we will assume that the fermion dispersion appearing in
Eq.~\eqref{eq:fermf} is quadratic, $\varepsilon (\ve{k}) = k^2/2m_F$,
where $\ve{k} =k_x$ if fermions, like bosons, move strictly in 1D,
whereas $\ve{k} = (k_x, k_y, k_z)$ if we instead consider the
mixed-dimensional case where fermions are free to move in all three
dimensions, while bosons still move strictly in 1D. Note that assuming
a quadratic isotropic dispersion for the fermions is expected to be a
good approximation in the small fermion density regime that we are
going to consider here, even if the true dispersion in the
mixed-dimensional case is anisotropic: In fact, when fermions start to
mix with bosons, they must necessarily occupy the lowest band energy
levels, for which the dispersion can be approximated as quadratic. The
anisotropy can be rescaled out when evaluating the fermion density
$\rho^{0}_{F}$ as a function of the Fermi wavevector $k_F$, and yields
to an overall prefactor relative to the isotropic result.

Considering the mean-field Hamiltonian derived in
Eq.~\eqref{eq:meanf}, the following step is to average over the
fermion operator density in the fermion Hamiltonian~\eqref{eq:fermf},
obtaining the Fermi-Dirac distribution function at zero temperature,
$\langle \hat{f}^{\dag}_{\ve{k}} \hat{f}^{\phantom{\dag}}_{\ve{k}}
\rangle = n_{\ve{k}} = \theta (k_F -k)$. The free energy density we
obtain this way will have a different dependence on the Fermi density
depending whether fermions move in 1D or 3D.

Let us start considering the case where fermions move in 1D, so that
the lineal Fermi density defined in Eq.~\eqref{eq:linde} is $\rho_{F}^0 =
N_F/NL= k_F/\pi$ and the free energy potential $g = \langle
\hat{H}^{mf}\rangle /LN$ reads as:
\begin{multline}
  g = \Frac{\pi^2}{6m_F} (\rho^{0}_{F})^3 -\left(\mu_F - g_{BF} A
  \rho^{0}_{B}\right)\rho^{0}_{F} +\\ \frac{(\rho_B^0)^3}{2m_B}
  e(\gamma) - \mu_B^{1D} \rho_B^0\; .
\label{eq:freeg}
\end{multline}
Note that now the free energy potential $g = g(\mu_B^{1D}, \mu_F,
\rho_B^0, \rho_F^0)$ depends on both the boson and fermion densities
as well as on the chemical potentials. This means that minimizing $g$
with respect to the fermion density, 
\begin{equation}
  f(\mu_B^{1D}, \mu_F, \rho_B^0) = \min_{\rho^{0}_{F}} g(\mu_B^{1D},
  \mu_F, \rho_B^0, \rho_F^0) \; ,
\label{eq:cond0}
\end{equation}
we retrieve the free energy potential $f(\mu_B^{1D}, \mu_F, \rho_B^0)$
considered in Eq.~\eqref{eq:freen}, which global minimum with respect
to the boson density gives the true thermodynamic grand-canonical free
energy density~\eqref{eq:minim}.

However, if we instead minimize $g(\mu_B^{1D}, \mu_F, \rho_B^0,
\rho_F^0)$ with respect to the boson density,
\begin{equation}
  h(\mu_B^{1D}, \mu_F, \rho_F^0) = \min_{\rho^{0}_{B}} g(\mu_B^{1D},
  \mu_F, \rho_B^0, \rho_F^0) \; ,
\label{eq:cond1}
\end{equation}
so that to eliminate $\rho^0_B$ in favor of $\rho^0_F$, $\mu_F$ and
$\mu_B^{1D}$, then the free energy potential $h(\mu_B^{1D}, \mu_F,
\rho_F^0)$ thus obtained can be used to study the phase transitions
where the number of Fermi surfaces changes from zero to one, by
expanding for small values of $\rho^0_F$. Because the dimensionless
function $e(\gamma)$ coming from the Bethe ansatz is determined
numerically, this plan of deriving the free energy potential
$h(\mu_B^{1D}, \mu_F, \rho_F^0)$ is not a trivial one. However, we can
carry on this procedure in two opposite limits, corresponding to high
($\gamma = m_B g_{BB}^{1D}/\rho_B^0 \ll 1$) and low ($\gamma \gg 1$)
density, respectively:
\begin{equation}
  e(\gamma) \simeq \begin{cases} \gamma &\gamma \ll 1 \\ e_{TG} =
    \frac{\pi^2}{3} &\gamma \gg 1 \; .
\end{cases}
\label{eq:lowde}
\end{equation}
Note that the last limit results in a contribution to the free energy
potential $h$ of $\pi^2 (\rho^{0}_{B})^3/6 m_B$, which is identical to
the kinetic energy of a free 1D Fermi gas.  This is the result of
fermionization in the Tonks-Girardeau limit~\cite{RMP_Cazalilla}.

Evaluating~\eqref{eq:cond1} by solving $\partial g/\partial
\rho^{0}_{B} = 0$ for $\rho^{0}_{B}$, in both cases of high and low
boson density, we obtain a series in terms of the Fermi density of
cubic form:
\begin{equation}
  h = h_0 + h_2 \rho^0_F + h_4 (\rho^0_F)^2 + h_6 (\rho^0_F)^3 +
  \cdots \; .
\label{eq:expa2}
\end{equation}
For high boson density the coefficients, for $\mu_B^{1D} > A g_{BF}
\rho^0_F$, which requires $\mu_B^{1D}$ and $g_{BF}$ to have the same
sign, are given by
\begin{align}
\label{eq:coef1}
  h_2 &= \frac{A g_{BF}}{g^{1D}_{BB}} \mu_B^{1D} - \mu_F\\
  h_4 &= - \frac{A^2 g^2_{BF}}{2g^{1D}_{BB}}\\
  h_6 &= \frac{ \pi^2}{6 m_F}\; . 
\end{align}
In the low boson density limit, instead, we obtain (also for
$\mu_B^{1D} > A g_{BF} \rho^0_F$) that the coefficients of the
expansion~\eqref{eq:expa2} are now given by
\begin{align}
\label{eq:coef2}
  h_2 &= \frac{A g_{BF}}{\pi} \sqrt{2 m_B \mu_B^{1D}} - \mu_F\\
  h_4 &=- \frac{A^2 g^2_{BF}\sqrt{m_B}}{2\pi \sqrt{2\mu_B^{1D}}} \\
  h_6 &= \frac{\pi^2}{6 m_F} - \frac{A^3 g_{BF}^3 \sqrt{m_B}}{12
    \sqrt{2} \pi {\mu_B^{1D}}^{3/2}}\; .
\end{align}
In this last case, note that, for $m_F \ll m_B$, as we have assumed in
the main text, the coefficient of the cubic term is expected to be
positive and large. In both cases, while $h_6>0$, the coefficient of
the square term is always negative, $h_4 < 0$, implying that the
transition cannot be continuous, rather it is first order. For small
fermion density, the transition has place for $h_2 > 0$ and $h_4 = -2
\sqrt{h_2 h_6}$, implying that, for large values of the coefficient
$h_6$, $h_2 \to 0$ and the transition is weakly first order, as we
have indeed observed numerically in many cases.

The $\rho^0_F$ expansion just carried on thus allows us to understand
the nature of the transition between the pure Bose gas and mixed
phases and deduce that it has to be first order in the strictly 1D
limit. We can merge this result with the one obtained in
Sec.~\ref{sec:mixdi}, where we were expanding for small boson density
(Tonks-Girardeau limit), and found that the transition between the
pure Fermi gas and the mixed phases is also first order. We can thus
conclude that in the 1D limit the transitions between the pure
boson/fermion phases and the mixed phase, must be all first order.

For comparison purposes, we can now consider the case of mixed
dimensionality, i.e., where fermions move in 3D. Now, the 3D Fermi
density is given by $N_F/\Omega = \rho^{0}_{F} d^{-2} = k_F^3/6\pi^2$,
where the volume is $\Omega = Nd^2 L$, $d$ is the spacing between the
tubes (cf. Fig~\ref{fig:schem}), and $\rho^{0}_{F}$ the lineal fermion
density. Thus the kinetic energy contribution to the free energy
potential $g = \langle \hat{H}^{mf}\rangle /LN$ now scales differently
than in the 1D limit:
\begin{multline}
  g = \Frac{d^2 ( 6 \pi^2 d^{-2} \rho^{0}_{F})^{5/3}}{20 \pi^2 m_F}
  -\left(\mu_F - g_{BF} A \rho^{0}_{B}\right)\rho^{0}_{F}
  +\\ \frac{(\rho_B^0)^3}{2m_B} e(\gamma) - \mu_B^{1D} \rho_B^0\; .
\label{eq:freg2}
\end{multline}
As before, we eliminate the boson density $\rho^0_B$ from the potential
$g$ by making use of~\eqref{eq:cond1} and then we expand for small
values of the Fermi density $\rho^0_F$. In this case we obtain:
\begin{equation}
  h = h_0 + h_2 \rho^0_F + h_{10/3} (\rho^0_F)^{5/3} + \cdots \; ,
\label{eq:expa3}
\end{equation}
where the coefficient $h_2$ is given by the
expressions~\eqref{eq:coef1} for the high boson density, while
by~\eqref{eq:coef2} for the low boson density limit. The coefficient
$h_{10/3}$ is in both cases instead given by:
\begin{equation}
  h_{10/3} = \Frac{d^{-4/3} ( 6 \pi^2 )^{5/3}}{20 \pi^2 m_F}\; .
\end{equation}
We can thus see that, while the coefficient $h_2$ can change sign
depending on the values of the system parameters, the coefficient of
the next order term, which scales like $(\rho^0_F)^{5/3}$ rather than
quadratically, is always positive $h_{10/3} > 0$. This means that the
position of the closest minimum to $\rho^0_F = 0$ is entirely
controlled by the sign of $h_2$, that is, the transition between the
pure boson and mixed phases must be continuous. By contrast, as we
have argued in Sec.~\ref{sec:mixdi}, the transition between the pure
fermion and the mixed phase is independent on the dimensionality where
fermions live and is always first order.

The main conclusion of the simple exercise carried on in this
appendix is that the different nature of the phase transitions where
the number of Fermi surfaces changes from zero to one between the
strictly 1D case and the mixed-dimensional case is a consequence of
the different scaling of the fermion kinetic energy in 1D and in
3D. Note that this result is also applicable to the transitions
between the pure fermion and mixed phases, as bosons in 1D at low
density behave as a free Fermi gas by virtue of fermionization.


\end{document}